\documentclass[12pt]{article}
\usepackage{amssymb,amsmath,amsthm,graphicx,ulem}

\pdfoutput=1

\usepackage{graphicx,subfigure}
\usepackage{epsfig}
\usepackage{amsmath}
\usepackage{amsfonts}
\usepackage{amssymb}
\usepackage[usenames]{color}
\usepackage[letterpaper,left=2.cm,right=2.cm,top=2.5cm,bottom=2.5cm]{geometry}
\usepackage{diagbox}
\usepackage{mathrsfs}

\newcommand{\beq}{\begin{equation}}
\newcommand{\eeq}{\end{equation}}
\newcommand{\be}{\begin{equation}}
\newcommand{\ee}{\end{equation}}
\newcommand{\beqa}{\begin{eqnarray}}
\newcommand{\eeqa}{\end{eqnarray}}
\newcommand{\beqar}{\begin{eqnarray*}}
\newcommand{\eeqar}{\end{eqnarray*}}
\newcommand{\bea}{\begin{eqnarray}}
\newcommand{\eea}{\end{eqnarray}}

\numberwithin{equation}{section}

\newcommand{\nn}\nonumber

\numberwithin{equation}{section}

\begin{document}

\allowdisplaybreaks

\normalem

\title{Weak cosmic censorship   in Born-Infeld electrodynamics and  bound on charge-to-mass ratio}

\author{ Tong-Tong Hu\footnote{hutt17@lzu.edu.cn}, Yan Song\footnote{songy18@lzu.edu.cn}, Shuo Sun\footnote{sunsh17@lzu.edu.cn},    \\ Hong-Bo Li\footnote{lihb2017@lzu.edu.cn},
 and  Yong-Qiang Wang\footnote{ yqwang@lzu.edu.cn, corresponding author}
\\ \\
   Research Center of Gravitation $\&$
 Institute of Theoretical Physics $\&$ \\
Key Laboratory for Magnetism and Magnetic of the Ministry of Education,\\ Lanzhou University, Lanzhou 730000, China
 \\
}

 \date{}

\maketitle

\begin{abstract}
\noindent We construct a  class of  counterexamples to cosmic censorship in four dimensional Einstein-Born-Infeld
theory with asymptotically anti-de Sitter boundary conditions, and investigate the effect of  the Born-Infeld
parameter $b$ in comparison with the counterpart of Einstein-Maxwell theory.
When a charged massive scalar field is included into the action, we find that this  class of counterexamples to cosmic
censorship  would be removed if the charge of  scalar fields is above the minimum value
of charge $q_{min}$. In particular,  the minimum value
of charge required to preserve cosmic censorship  increases with the increasing of  Born-Infeld parameter. Meanwhile,
we also show   the  lower bounds on charge-to-mass ratio with the different values of Born-Infeld parameter.
\end{abstract}

\newpage


\tableofcontents

\section{Introduction}
The study of  black hole singularities has been an interesting subject since the original
work on the weak cosmic censorship conjecture (WCCC) \cite{Penrose:1969pc}, which states  naked singularities  arising in the solutions of Einstein's equations \cite{Hawking:1976ra}  must be  hidden within event horizons of black hole, and therefore cannot be observed from future null infinity.
Although there existed many works of testing the validity of WCCC, lots of counterexamples to cosmic censorship have been found.
For example, the collapse of a massive matter cloud with regular initial data  results in the formation of a naked singularity \cite{Joshi:2012mk}.
It is an interesting  question to ask  whether there exists a region of arbitrarily large curvature that is observed
to distant observers.
In higher dimensions,
the event horizon of a class of black holes is not necessarily the case of topologically spherical.  Due to the Gregory-Laflamme instability \cite{Gregory:1993vy}
,  black holes will be unstable under gravitational perturbations  and
produce a naked singularity, which implies the violation of cosmic censorship\cite{Hubeny:2002xn,Lehner:2010pn,Santos:2015iua}.
Analysing the non-linear evolution
of black holes \cite{Figueras:2015hkb,Figueras:2017zwa}, one found that a naked singularity should
form in finite  time. In four dimensions,
by investigating the model of Einstein gravity coupling to a Maxwell field with a negative cosmological
constant \cite{Horowitz:2016ezu,Crisford:2017zpi}, one could obtain  a  class of
counterexamples to cosmic censorship  with asymptotically anti-de Sitter
(AdS) boundary. Instead
of adding a Maxwell field, the authors in Ref. \cite{Crisford:2018qkz} constructed smooth stationary solutions with differential rotation to the boundary metric, which  provides a possible vacuum counterexamples to weak cosmic censorship in AdS spacetime.

Recently, by introducing a charged scalar field  to the
Einstein-Maxwell solutions \cite{Crisford:2017gsb} , the authors found that when the charge of the scalar field was  sufficient large,
the static solution of Einstein-Maxwell action would
become unstable and a new stable  solution could appear with nontrivial charged scalar field, which is   analogous to the instability of a charged black hole to develop
scalar condensation in the study of  holographic superconductor \cite{Gubser:2008px,Hartnoll:2008vx,Cai:2015cya}.
Meanwhile,
there exists a  minimum value
of charge required to  remove the counterexamples and preserve cosmic censorship.
It is surprising that
 the  minimum value
of charge appears to agree precisely with that proposed in
the weak gravity conjecture \cite{ArkaniHamed:2006dz}, which states that any consistent quantum theory of gravity must have a stable
state whose
charge-to-mass ratio is equal to or larger than that of an extremal black hole. In the case of  Einstein-Maxwell model, the bound of $q/m$ is equal to $1$.
To further test  the weak gravity - cosmic
censorship connection, the author in \cite{Horowitz:2019eum}
 also studied  the static solutions   in Einstein theory with a dilaton field and  the multi-charged scalar field case.

Considering that Maxwell theory is only the theory of linear electrodynamics, we would like to know  whether or not there exists  a  class of  counterexamples to cosmic censorship and the charge-to-mass ratio bound in four dimensional Einstein-nonlinear electrodynamics
theory.
  The Born-Infeld electrodynamics \cite{IN-Born:1933gh,IN-Born:193444,IN-Born:1934gh} is  a nonlinear generalization of the Maxwell's theory,  which can remove the divergence of self-energy of a point-like charge in Maxwell electrodynamics. Moreover, the Lagrangian of Born-Infeld can arise from the low-energy effective theory describing electromagnetism \cite{Fradkin:1985qd,Leigh:1989jq}. Many  works   on  the black holes solutions in Einstein-Born-Infeld theory have been studied in \cite{gsp,Cataldo:1999wr,IN-Fernando:2003tz,IN-Dey:2004yt,IN-Cai:2004eh,IN-Li:2016nll}.

    In the present paper,  we would like to
     construct a  class of  counterexamples to cosmic censorship in four dimensional Einstein-Born-Infeld
theory with asymptotically anti-de Sitter boundary. Furthermore,  comparing with the counterpart of Einstein-Maxwell theory,  we investigate  the counterexample  of Einstein-Born-Infeld theory  for several values of  Born-Infeld
parameter $b$.
Besides, introducing a charged massive scalar field into Einstein-Born-Infeld theory, we find that this  class of counterexamples to cosmic
censorship  would be removed, and
 we study the value of  the  lower bound on the scalar field charge required to preserve cosmic censorship
 in the case of  Born-Infeld action.

 The paper is organized as follows. In Sec. \ref{sec2}, we introduce the model of Einstein-Born-Infeld coupling to a complex, charged scalar field
   and the numerical  DeTurck method. In Sec. \ref{subsubsec:boundary},  we explore the
ansatz of metric and matter field,  and analyze the boundary conditions. Numerical results of a class of  counterexamples to cosmic censorship and
    static solutions  with charged scalar condensation are  shown  in Sec. \ref{nrss}.  The conclusion and discussion are given in the
  last section.

\section{Set up}\label{sec2}
Let us begin with the action of  the Born-Infeld field and a charged complex scalar field in the four-dimensional Einstein gravity spacetime with a negative cosmological constant, which is written as
 \begin{equation}
 S = \frac{1}{16\pi G}\int \mathrm{d}^4 x\,\sqrt{-g}\left[R+\frac{6}{L^2}+\mathcal{L}_{BI}-4\,(\mathcal{D}_a \Phi)(\mathcal{D}^a \Phi)^\dagger-4\,m^2\Phi \Phi^\dagger\right],\,
 \label{eq:action}
 \end{equation}
where $\mathcal{L}_{BI}=\frac{4}{b}(1-\sqrt{1+\frac{bF}{2}})$  with the
field strength of the U(1) gauge field $F=F_{ab}F^{ab}$,  and $L$ is the radius of
asymptotic AdS spacetime.
The constants $ m$ and $q$ represent the
mass and the charge of the complex scalar field ¦×, respectively.
The constant b is the Born-Infeld parameter and the Born-Infeld  field will reduce to the
Maxwell case when $b\rightarrow 0$.	Where $\mathcal{D}_a=\nabla_a-i q A_a$ is the gauge covariant derivative with respect to $A_a$. Note that the values of $m^2$ must satisfy the
Breitenlohner-Freedman (BF) bound $m^2 \geq -9/4$ \cite{B_F} for the (3+1)-dimensional
spacetime.
	\par
	The motion equations can derived from Eqs. (\ref{eq:action})
	\begin{subequations}
		\begin{align}
		&R_{ab} +\frac{3}{L^2}g_{ab}= T_{ab},\label{eq:einstein}
		\\
		&\nabla_a(-\mathscr{F}F^{a}_{\phantom{a}b})=i\,q\,\left[(\mathcal{D}_b \Phi)\Phi^\dagger-(\mathcal{D}_b \Phi)^\dagger\Phi\right]\,,
		\\
		&\mathcal{D}_a\mathcal{D}^a \Phi = m^2 \Phi\,,
		\label{eq:scalar}
		\end{align}
		\label{eqs:motion}
	\end{subequations}
with the energy-momentum tensor of matter field
\begin{equation*}
T_{ab}=2\left(-\mathscr{F}F_{a}^{\phantom{a}c}F_{bc}+\frac{g_{ab}}{4}\mathcal{L}_{BI}\right)+2(\mathcal{D}_a \Phi) (\mathcal{D}_b \Phi)^\dagger+2(\mathcal{D}_a \Phi)^\dagger (\mathcal{D}_b \Phi)+2\,m^2 g_{ab}\Phi\Phi^\dagger\,,
\end{equation*}
	where $\mathscr{F}\equiv\frac{\partial \mathcal{L}_{BI}}{\partial F}$, and it equals to -1 in Maxwell condition.
	\par

If the  complex, massive  scalar field $\psi $ vanishes, the solution of Einstein equations (\ref{eqs:motion}),  which can describe  the asymptotically spherically
   black
hole with charge, is the well-known Born-Infeld AdS black hole.
In terms of spherical coordinates,
		 the  metric  of Born-Infeld AdS black hole is the following form
	\begin{equation}
		ds^2=-f(r)dt^2+f^{-1}(r)dr^2+r^2(d\theta^2+\sin^2\theta d\phi^2),
		\end{equation}
with
		\begin{equation}
		f(r)=1-\frac{2M}{r}+\frac{r^2}{L^2}+\frac{4Q^2 {}_2F_1(\frac{1}{4},\frac{1}{2};\frac{5}{4};-\frac{bQ^2}{r^4})}{3r^2}+\frac{2r^2}{3b}\left(1-\sqrt{\frac{bQ^2}{r^4}+1}\right),
		\end{equation}
where  $M$ and $Q$ are the ADM mass and  the electric charge  of BI  AdS black hole, respectively,   and ${}_2F_1$ is a hypergeometric function \cite{hugeome}. The gauge potential is

	\begin{equation}
    A= -\frac{Q {}_2F_1(\frac{1}{4},\frac{1}{2};\frac{5}{4};-\frac{bQ^2}{r^4})}{r} dt.
	\end{equation}
	\par

When   there exists a non-trivial  configuration of the complex scalar field,  it is obvious that we should solve the equations of motion (\ref{eqs:motion})
numerically instead of seeking the analytical solutions.
We will use DeTurk method to solve these equations, which provides a good tool for solving Einstein equations in these papers. By adding a gauge fixing term to Einstein equation (\ref{eq:einstein}), we could obtain a set of  elliptic equations, which are  known as  Einstein-DeTurk equation
	\begin{equation}
	R_{ab}+\frac{3}{L^2}g_{ab}-\nabla_{(a}\xi_{b)}=T_{ab},
	\end{equation}
where $\xi^a=g^{bc}(\Gamma^{a}_{bc}[g]-\Gamma^{a}_{bc}[\tilde{g}])$ is the Levi-Civita connection associated with a reference metric $\tilde{g}$,
which should be choose to be as same boundary and horizon structure as $g$.

\section{\label{subsubsec:boundary}Ansatz and boundary conditions}
In order to construct static, axisymmetric solutions with a timelike Killing vector  and
an axisymmetric Killing vector, we also adopt the same axisymmetric metric as that in Refs. \cite{Horowitz:2014gva,Horowitz:2016ezu,Crisford:2017gsb,Horowitz:2019eum} with the following ansatz
\begin{multline}
d s^2=\frac{L^2}{\left(1-x^2\right)^2}\Bigg[-\frac{\left(1-y^2\right)^2\,U_1\,d t^2}{y^2 \left(2-y^2\right)}+\frac{4\,U_4}{2-x^2}\left(d x+\frac{U_3}{1-y^2}d y \right)^2\\
+\frac{4\,U_2\,d y^2}{y^2 \left(1-y^2\right)^2\left(2-y^2\right)^2}+x^2 \left(2-x^2\right)\,U_5\,d\phi^2\Bigg]\,,
\label{ansatz}
\end{multline}
where the functions $U_i ~ ( i= 1,2,3,4,5)$  depend on the variables  $x$  and  $y$.
Both of variables take values in $[0, 1]$.
When $U_1 = U_2 = U_4 = U_5 = 1$ and
$U_3 = 0$, the metric (\ref{ansatz}) can reduce to
\begin{equation}
ds^2=\frac{L^2}{\left(1-x^2\right)^2}\left[-\frac{\left(1-y^2\right)^2 d t^2}{y^2 \left(2-y^2\right)}+\frac{4\,d x^2}{2-x^2}+\frac{4\,d y^2}{y^2 \left(1-y^2\right)^2\left(2-y^2\right)^2}+x^2 \left(2-x^2\right)d\phi^2\right]\,.
\label{eq:purenew}
\end{equation}
 When we  use a new coordinate system,
\begin{subequations}
\begin{align}
&z=\frac{y \sqrt{2-y^2}}{1-y^2}(1-x^2)\,,
\\
&r=\frac{y \sqrt{2-y^2}}{1-y^2}x\sqrt{2-x^2}\,,
\end{align}\label{nn}
\end{subequations}
the line element (\ref{eq:purenew}) in new coordinates becomes
	\begin{equation}
ds^2=\frac{L^2}{z^2}\left[-d t^2+d r^2+r^2 d\phi^2+d z^2\right],
\label{eq:pure}
\end{equation}
which is just the  pure Anti-de Sitter spacetime in Poincar\'e coordinates.
The Poincar\'e horizon is now located at y = 1, and the axis of rotation is located at x = 0.
The conformal boundary is located at x = 1, and y = 0 denotes the intersection of the
conformal boundary with the axis of symmetry.  We can choose the reference metric $\tilde{g}$  given by the line element of pure Anti-de Sitter spacetime (\ref{eq:purenew}).

Considering the above metric (\ref{ansatz}), an ansatz of matter fields should be described as
below
\begin{equation}
 A=L U_6 dt,\;\;\;\;\;\;\; \Phi = \left(1-x^2\right)^\Delta y^\Delta \left(2-y^2\right)^{\frac{\Delta}{2}}U_7\,
\label{scalr}
\end{equation}
with
\begin{equation}
 \Delta\equiv 3/2+\sqrt{9/4+m^2},
\label{scaler1}
\end{equation}
where the function $U_6$  and $U_7$ are the function of x and y.
According to AdS/CFT duality, we take  $\langle \mathcal{O}_2\rangle=(1-y^2)^2 U_7$ to describe the scalar condensation.
To simplify, we only choose $m^2 = -2$ in our paper.

 Next, we will discuss the boundary conditions. At comformal boundary, located at x=1, the metric must reduce to pure AdS spacetime, so we must take
 \begin{equation}\label{boundary1}
  U_1(1,y)=U_2(1,y)=U_4(1,y)=U_5(1,y)=1,  \;\;\;\;\; U_3(1,y)=0,\;\;\;\;\;
 \end{equation}
and the boundary of matter fields
\begin{equation}\label{boundary2}
\partial_x U_7(1,y)=0, \;\;\;\;\; U_6(1,y)=a(1-y^2)^n,\;\;\;n=2,4,6, \cdots,
 \end{equation}
where $a$ is the amplitude. At $x=0$, we could impose the  Dirichlet boundary conditions on
 $U_3(1,y)=0$
and
 \begin{equation}\label{dada1}
\partial_x U_1(0,y)=\partial_x U_2(0,y)=\partial_x U_4(0,y)=\partial_x U_5(0,y)=\partial_x U_6(0,y)=\partial_x U_7(0,y)=0,
 \end{equation}
Neumann boundary conditions on the other functions.
Moreover, the asymptotic behavior of the equations of motion near $x = 0$ gives the condition $U_4(0,y) = U_5(0,y)$.

 At y=0, i.e. the intersection of the conformal boundary with the axis of symmetry, we set the boundary conditons
  \begin{equation}\label{dada2}
  U_1(x,0)=U_2(x,0)=U_4(x,0)=U_5(x,0)=1, \;\; U_3(x,0)=\partial_y U_7(x,0)=0, \;\;  U_6(x,0)=a,
  \end{equation}
and at Poincar\'e horizon y=1, Dirichlet boundary conditions are imposed on
 \begin{equation}\label{dada2}
  U_1(x,1)=U_2(x,1)=U_4(x,1)=U_5(x,1)=1, \;\; U_3(x,1)= U_6(x,1)= U_7(x,1)=0.
  \end{equation}

\section{Numerical results}\label{nrss}
In this section, we will numerically solve the coupled system of nonlinear partial differential equations (\ref{eqs:motion}) with the ansatzs (\ref{ansatz}) and (\ref{scalr}).
 We  employ finite element methods in the integration region  $0\leq x\leq 1$ and $0\leq y \leq 1$ defined on non-uniform grids, allowing the grids to be more finer   grid points near the boundary  of  $y=0$ and  $y=1$. Our iterative process is the Newton-Raphson  method. The relative  error  for the numerical solutions in this work is estimated to be below $10^{-5}$. In order to keep good agreement with the
aforementioned  error, the grid
size has to be increased and typically a $120\times200$ to
$120\times 300$ grid was used.

 We would  study the following two  cases with scalar field $\Phi=0$ and  $\Phi\neq0$, respectively.
For convenience, in the following results, we will set $L = 1$.
\subsection{\label{subsubsec:without}With charged scalar field  $\Phi=0$}
\begin{figure}[h!]
\begin{center}
	\includegraphics[height=.34\textheight,width=.38\textheight, angle =0]{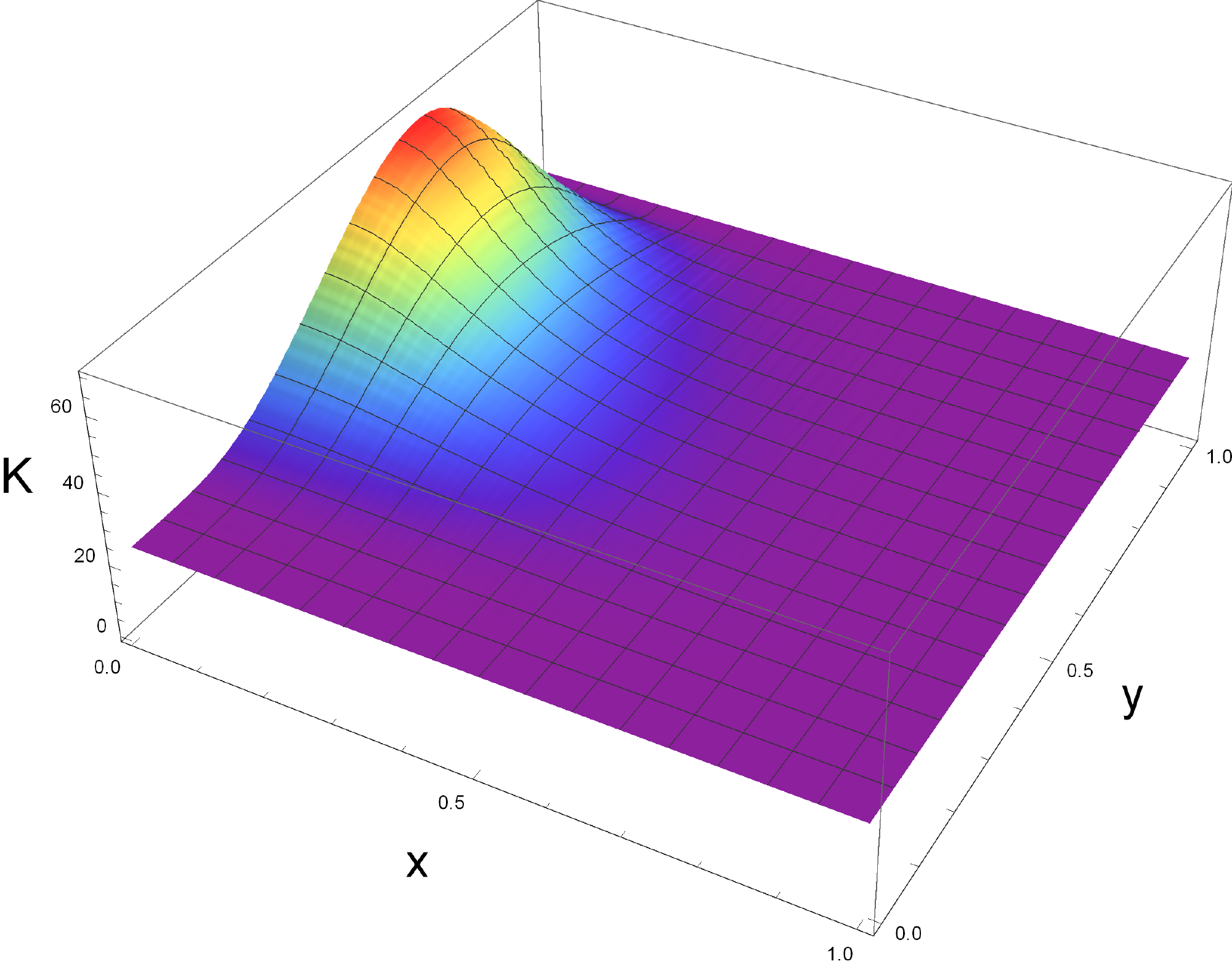}
		\includegraphics[height=.36\textheight,width=.38\textheight, angle =0]{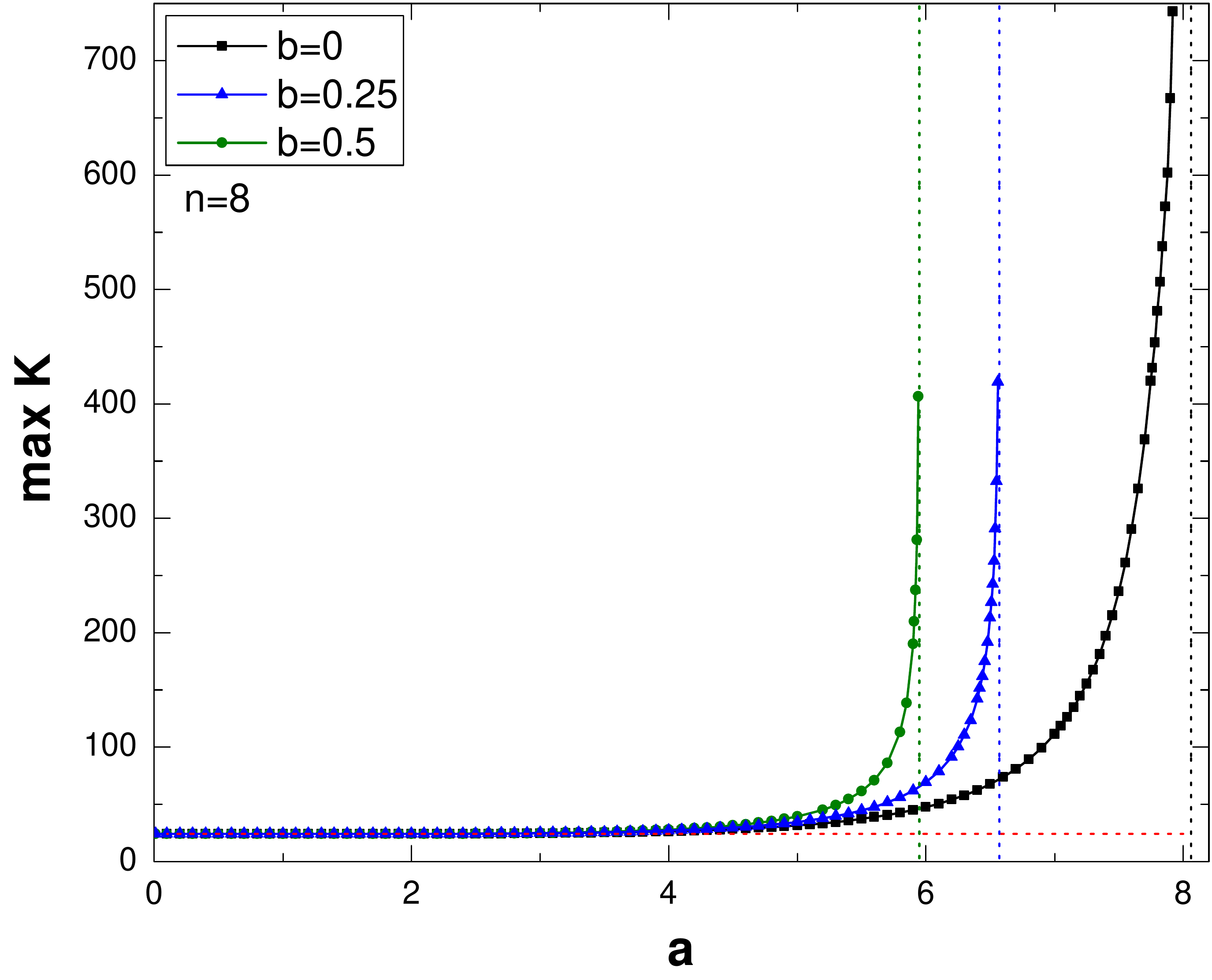}
\end{center}
	\caption{\textit{Left}: The distribution of  Kretschmann scalar  as a function of  $x$ and $y$ coordinate  with  $b=0.25$, $n=8$ and  $a=6$.                                                                                                                                                                                               \textit{Right}: The maximum of the Kretschmann scalar as a function of  $a$  with the fixed Born-Infeld
parameters   $b=0,~0.25,~0.5$, represented by the black,  blue  and  cyan lines, respectively.  The  vertical   black, blue  and cyan  dashed gridlines indicate   $a_{max}=8.06$, $a_{max}=6.56$ and $a_{max}=5.94$, respectively,
and  the horizonal  red  line is  the value of $K=24$ for AdS$_4$ spacetime.}
	\label{kstability}
\end{figure}	
In this subsection,  we will analyze the  solutions with  charged scalar field  $\Phi=0$ and present the evidence
for counterexample to cosmic censorship.
When one obtains a solution of Einstein equations, it is  important to investigate whether or not the spacetime of solution is
regular. In general, one of the most useful ways is to check for the
finiteness of the Kretschmann scalar,  which  is also called Riemann tensor
squared and written as
\begin{equation}
  K=R_{\alpha\beta\gamma\delta}R^{\alpha\beta\gamma\delta},
\label{eq:KRECUR}
\end{equation}
where $R_{\alpha\beta\gamma\delta}$ is the Riemann curvature tensor. Therefore,  Kretschmann scalar is a sum of squares of tensor components and  a quadratic invariant.

Numerical results are presented in Fig~\ref{kstability}. In the left panel we present the distribution of  Kretschmann scalar $K$ as a function of  $x$ and $y$ coordinate  with  $b= 0.25$, $n=8 $ and  $a= 6$. and it is obvious that the spacetime is not flat, and
the value of the Kretschmann scalar in the purple area is about to $24/L^4$,  which means  one  recover the result
of  $AdS_4$ spacetime. In addition, the maximum of the Kretschmann scalar  appears at  the boundary $x=0$.
In the right panel,  we exhibit   the maximum of the Kretschmann scalar $K$
versus the amplitude  $a$  with the fixed Born-Infeld
parameters   $b=0,~0.25,~0.5$, represented by the black,  blue and green lines, respectively, and 
the vertical black, blue and cyan dashed gridlines indicate $a_{max}= 8.06$,  6.56
and 5.94, respectively, and the horizonal red line is the value of $K = 24$ for $AdS_4$ spacetime.
From the figure, we could find that when the Born-Infeld
parameters $b\neq0$, there still exists the growth of the maximum of Kretschmann scalar with the increasing of $a$, which indicates the formation of a curvature singularity similar to the case of Maxwell action in \cite{Crisford:2017gsb}.

	When the amplitude reaches to a maximum value, there will appear a singularity, which means producing arbitrarily large curature in spacetime. This maximum is denpended on both $n$ and $b$. We show our result in the following table.
	\begin{table}[!htbp]
		\centering
		\begin{tabular}{|c|c|c|c|c|}
			\hline
       \diagbox{b}{$a_{max}$}{n} &$4$&$6$&$8$&$10$\\
            \hline
			$0.00$&4.95&6.53&8.06&9.13\\
			\hline
			$0.25$&4.22&5.51&6.56&7.45\\
			\hline
			$0.5$&3.82&4.99&5.94&6.76\\
			\hline
			$1$&3.35&4.36&5.20&5.91\\
			\hline
		\end{tabular}
\caption{Maximum amplitude $a_{max}$ for several values of $n$ and $b$.}\label{table1}
	\end{table}

Form  Table \ref{table1}, we could see that  the maximum amplitude $a_{max}$ increases with the  increasing of $n$, meanwhile, it decreases  with the  increasing of
the Born-Infeld parameter $b$. When Maxwell field   is extended to   Born-Infeld model, the static solutions could still become singular at a finite value of $a$.

\subsection{\label{subsubsec:turck}With  charged scalar field $\Phi\neq0$}
In the last subsection, we obtain a family of Einstein-Maxwell solutions with  charged scalar field $\Phi=0$,
we would like to know whether or not there exist the charge static solutions with no-vanished scalar field, which is analogous to  the case of holographic superconductor studied in \cite{ Gubser:2008px,Basu:2008st,Hartnoll:2008kx,Cai:2015cya}. When fixed the parameter $b$ and $a$, there is a critical value of $q$,   below which the scalar field vanishes and the solution is the simple planar. Above this value, the  charge static solution becomes unstable to develope scalar hair. Before numerically solving the full  dynamic
equations of motion including scalar field in Eqs. (\ref{eqs:motion}), we would follow the same method as in \cite{Horowitz:2013jaa,Crisford:2017gsb}, and find the critical value of $q$.
First, we solve the time-independent scalar field equations at a fixed background.
\begin{equation}
(\nabla_a \nabla^a -m^2)\Phi = q^2\,A_a A^a\,\Phi\,,
\label{eq:eige}
\end{equation}
which one could recognize as an eigenvalue problem   with the eigenfunction $\Phi$ and eigenvalue $q^2$. We will use $q_{min}$ to denote the smallest eigenvalue, which is the critical charge and needed as a function of the amplitude $a$ for a zero-mode.
\begin{figure}[h!]
\begin{center}
		\includegraphics[height=.36\textheight,width=.38\textheight, angle =0]{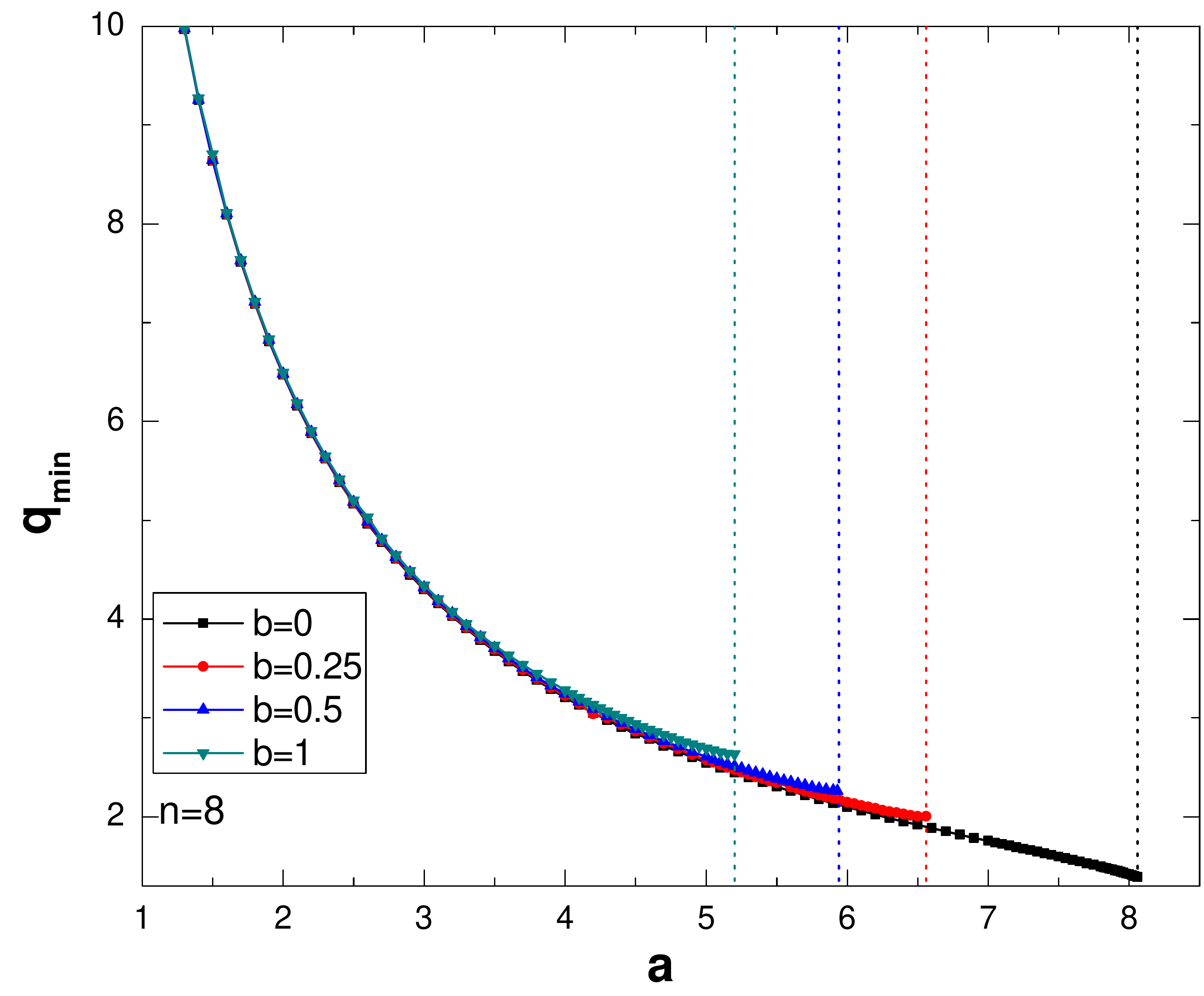}
		\includegraphics[height=.36\textheight,width=.38\textheight, angle =0]{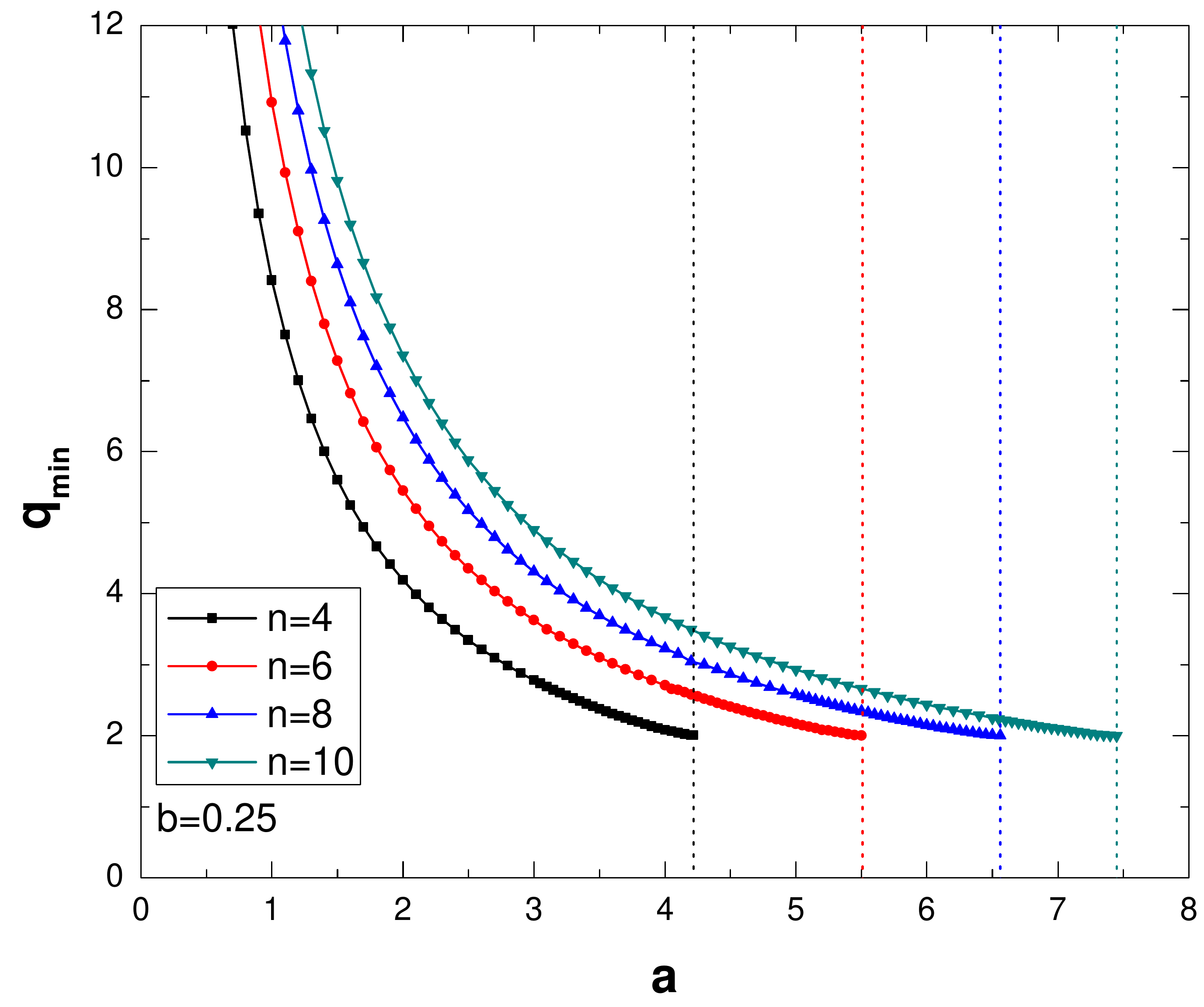}
\end{center}
	\caption{The critical charge $q_{min}$  as a function of the amplitude $a$.
\textit{Left}: For the fixed $n=8$, the curves from top to bottom correspond to $b = 1, ~0.5, ~0.25, ~0$,
respectively. \textit{Right}: For the fixed $b=0.25$, the curves from top to bottom correspond to $n = 10, ~8, ~6, ~4$,
respectively.}
	\label{stability1}
\end{figure}	

The results are shown in  Fig~\ref{stability1}.
In left panel, with the fixed $n=8$,  we plot  critical
charge   $q_{min}$   as a function of the amplitude  $a$  for the corresponding  values of  $b=1, 0.5$, $0.25$ and 0, represented by  cyan,  blue,  red  and black lines, respectively.
 The  vertical    dashed gridlines indicate the  maximum amplitude $a_{max}=8.06$ (black), $6.56$ (red), $5.94$ (blue) and $5.20$ (cyan), respectively, which corresponds to the values of $n=8$  shown in Table \ref{table1}.
For each curve, the critical charge $q_{min}$ decreases with the increasing
amplitude $a$.
We see that for the fixed amplitude $a$, the critical charge $q_{min}$ increases with increasing Born-Infeld parameter $b$. Furthermore,  when the amplitude $a$ decreases,  the critical charge $q_{min}$  tends to the same value for  several values of $b$.

 In the right of Fig~\ref{stability1}, with the fixed $b=0.25$,
 we plot  critical
charge   $q_{min}$   as a function of the amplitude  $a$  for the corresponding  values of  $n=10, ~8, ~6$ and $4$, represented by  cyan,  blue,  red  and black lines, respectively. For each curve, the critical charge $q_{min}$ decreases with the increasing
amplitude $a$, which are similar to the behavior in the left panel.
For the fixed amplitude $a$, the critical charge $q_{min}$ increases with the increasing of $n$.

\begin{figure}[h!]
	\begin{center}
		\includegraphics[height=.36\textheight,width=.38\textheight, angle =0]{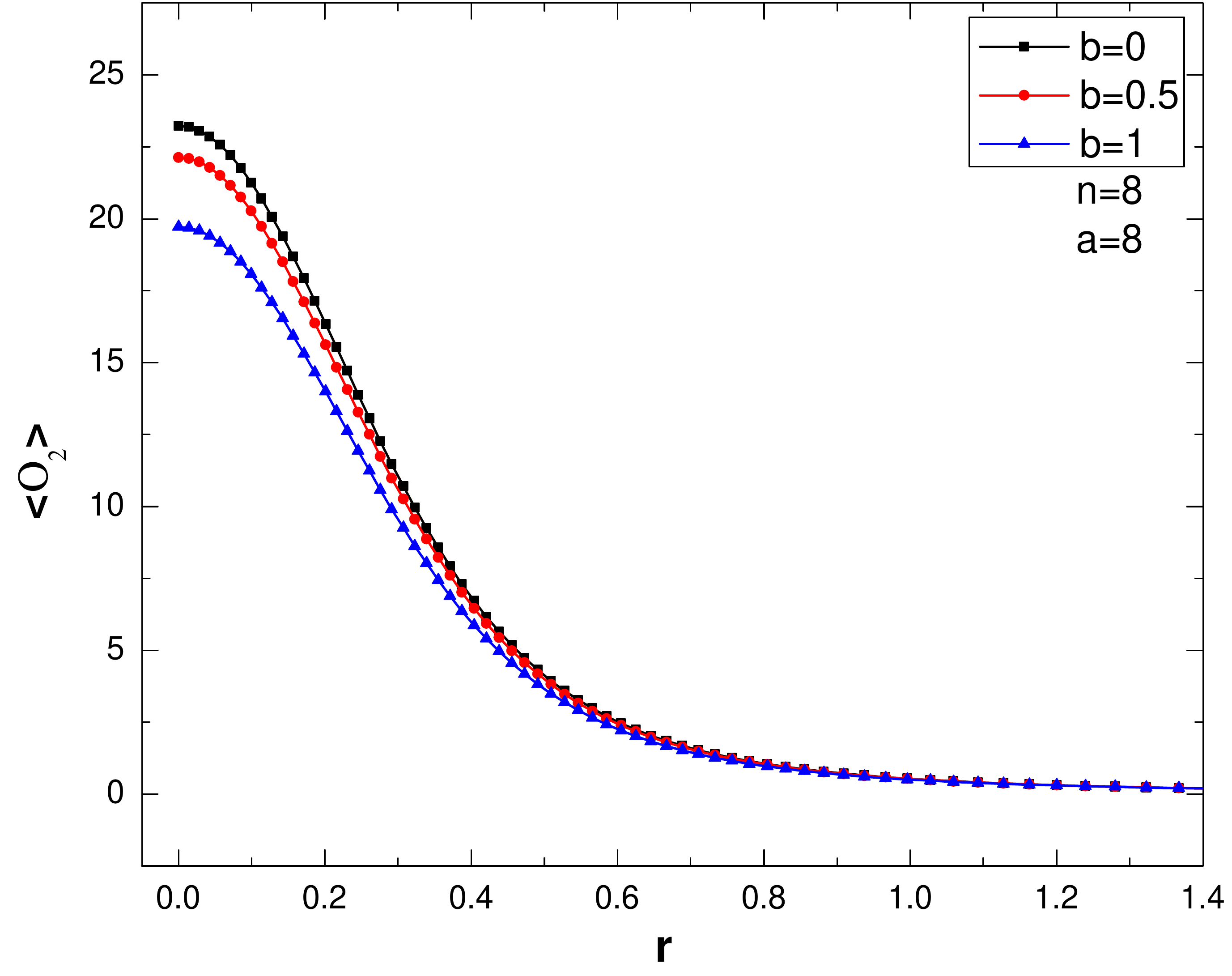}
		\includegraphics[height=.36\textheight,width=.38\textheight, angle =0]{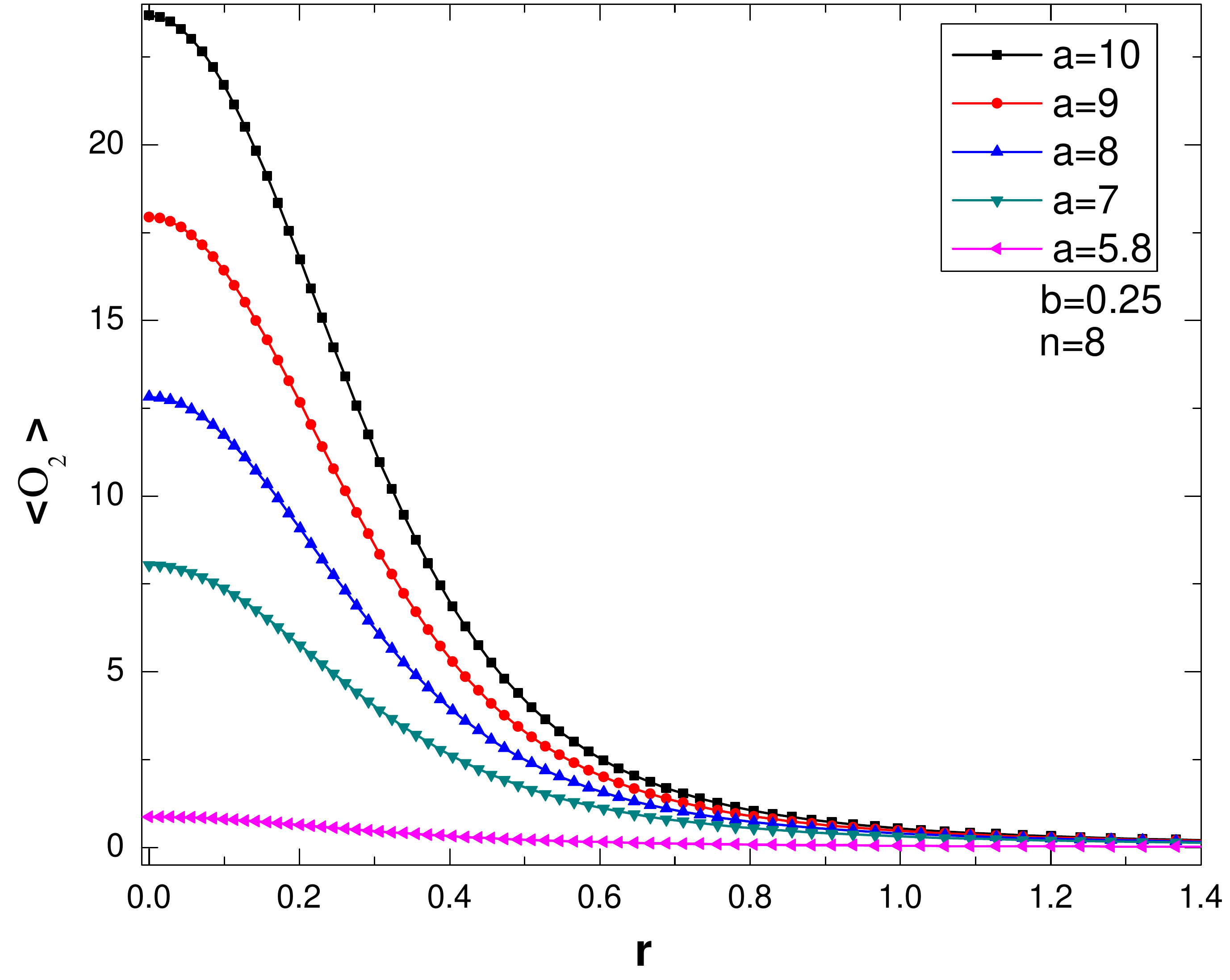}
	\end{center}
	\caption{
The condensation as a function of  radial coordinate $r$ at conformal boundary with different $b$ (left) and  $n$ (right). At both of these two figures we take the value of  $n=8$.}
	\label{condensation}
\end{figure}
Next, we could solve the coupled Einstein-Born-Infeld-scalar equations (\ref{eqs:motion}) to find the solutions for the above critical charge $q<q_{min}$.
 We show the profiles of the expectation
value for the operator dual to $\Phi$ as a function of the boundary radial coordinate $r$ in Fig.~\ref{condensation}. In the left panel,  with the fixed $n=8$ and $q=2.7$, the curves from top to bottom correspond to $b =0,~0.5,  ~1$, represented by
black, red and blue
 lines, respectively. For each curve, the scalar condensation $\langle \mathcal{O}_2\rangle$ decreases with the increasing
amplitude $r$.
  When the amplitude $r$ increases,  the scalar condensation   tends to zero value for several values of $b$.
Moreover,  at the origin $r = 0$, there exists the largest value of  the scalar condensation.
Furthermore, for the fixed amplitude $r$, the scalar condensation $\langle \mathcal{O}_2\rangle$ decreases with the increasing value of Born-Infeld parameter $b$.

 In the right of Fig.~\ref{condensation}, with the fixed $b=0.25$ and  $q=2.22$,
 we plot  the scalar condensation $\langle \mathcal{O}_2\rangle$ as a function of  $r$   for   several values of $a$.
The curves from top to bottom correspond to $a =10,~9,~8,~7$ and  $5.8$, represented by
black, red, blue, cyan and purple
 lines, respectively.
  For each curve, the scalar condensation decreases with the increasing
 $r$, which is similar to the behavior in the left panel.
For the fixed amplitude $b$, the scalar condensation $\langle \mathcal{O}_2\rangle$ increases with the increasing value of the  parameter $a$.

\begin{figure}
\begin{center}		
\includegraphics[width=0.49\textwidth,height=0.35\textheight]{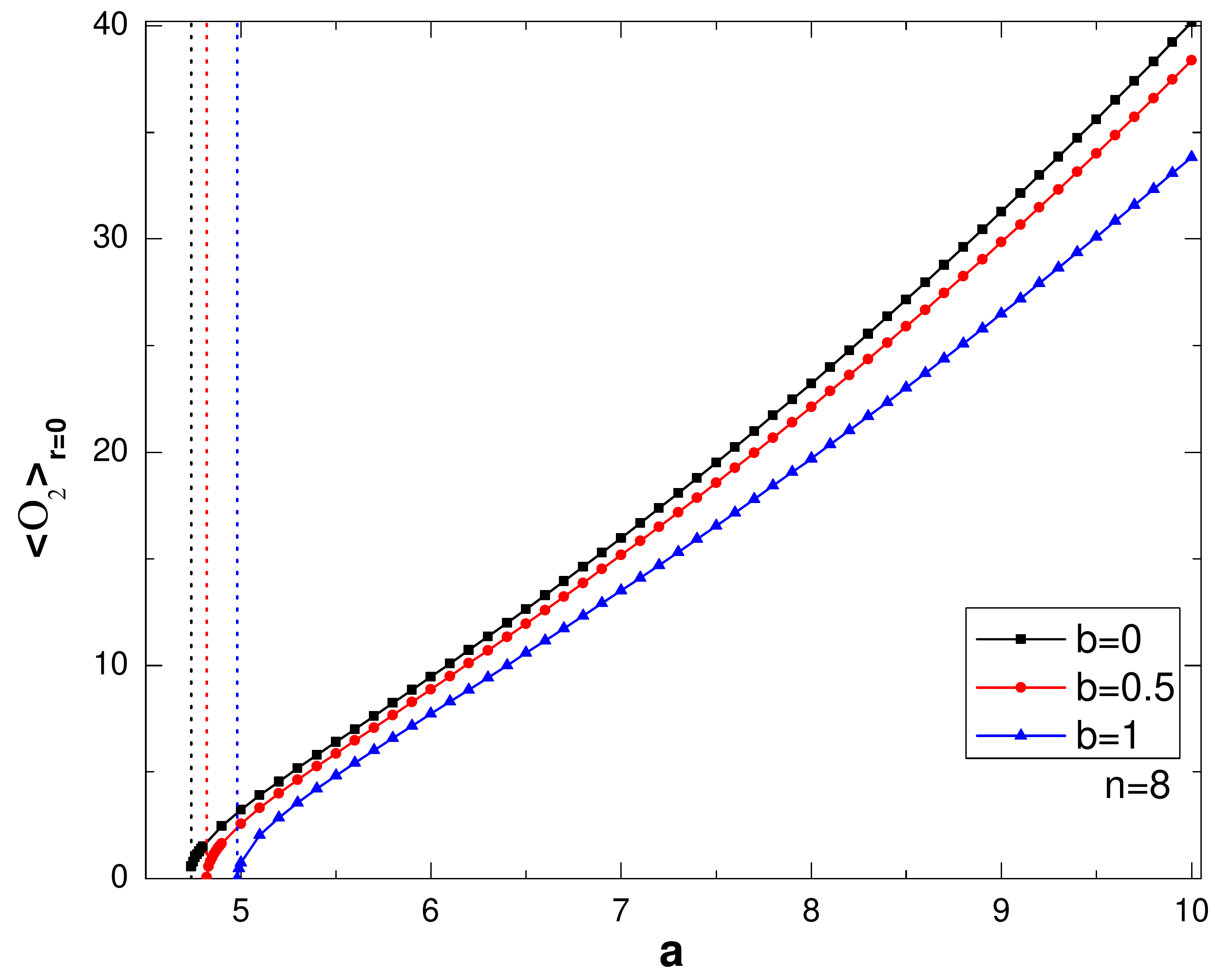}
\includegraphics[height=.35\textheight,width=.38\textheight, angle =0]{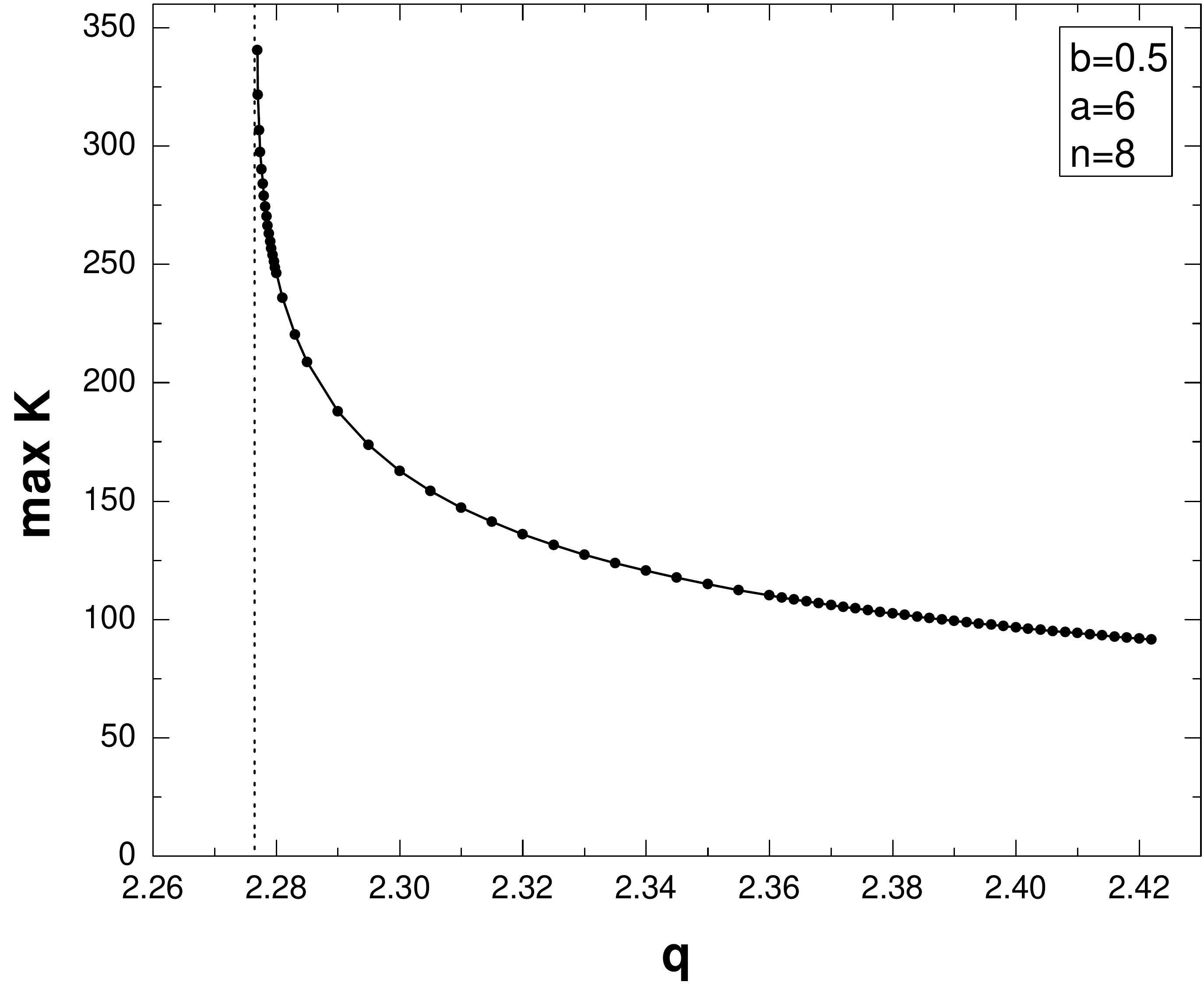}
	\end{center}
	\caption{\textit{Left}: The condensation against $a$ with different values of $b$ at r=0. \textit{Right}: The typical maximal value of  Kretschmann scalar $K$ as a function of the charge $q$
 with  $b=0.5$, $n=8$ and  $a=6$. }
	\label{r=0condensation}
\end{figure}

In order to further understand how increasing the amplitude $a$ affects the maximum of the condensation with the different values of $b$,
we show in the left panel of Fig.~(\ref{r=0condensation}) the condensate $\langle \mathcal{O}_2\rangle$ at $r=0$ as a function of the  amplitude $a$  with the fixed parameter $n=8$, $q=2.7$.
From top to bottom, the curves correspond to the values of $b=0$, $b=0.5$, and $b=1$, respectively.  In addition,  the  vertical  dashed gridlines  indicate the critical  amplitude $a_c=4.74$ (black), $a_c=4.82$ (red) and $a_c=4.98$ (blue), respectively.
 As one can see that there exists a critical  amplitude $a_c$
 above which the condensate appears, then rises  as the system is imposed with much larger amplitudes $a$. This behaviour of $b\neq0$ is qualitatively
similar to that  in Maxwell theory with $b=0$.

To investigate the connection between cosmic censorship and the weak gravity conjecture
 in Born-Infeld electrodynamics, we now study whether or not  there exists a critical $q_w$, above which weak
 cosmic censorship could still be preserved. It is noteworthy that below the critical $q_w$ there still exists  the scalar
field condensate before reaching a singularity. Because now the weak gravity conjecture in Born-Infeld electrodynamics is still not clear,
one could not determine whether the minimal value of charge-to-mass ratio is equal to  weak gravity conjecture.  However,  we could still obtain a  class of  static solutions with charged scalar condensate in the case of Born-Infeld electrodynamic,  which would prevent the violation of the weak cosmic censorship  if sufficient charged particles
 were present.

After obtaining  the numerical solution with charged scalar condensate,  we present in the right panel of Fig.~\ref{r=0condensation} the typical maximal value of  Kretschmann scalar $K$ as a function of the charge $q$
 with  $b=0.5$, $n=8$ and  $a=6$, and the   vertical  black  dashed gridlines indicate the $q_c=2.276$ minimal value.
From this plot, we can see that
the full nonlinear solutions with the scalar condensate are in  the range of  $ q > q_{c}$.
It is obvious that  there exists the growth of the maximum of Kretschmann scalar with the decreasing of $q$, which indicates the formation of a curvature singularity at some critical value of $q = q_c$, and is similar to the case of Maxwell case.

\begin{figure}
\begin{center}		\includegraphics[width=0.6\textwidth,height=0.35\textheight]{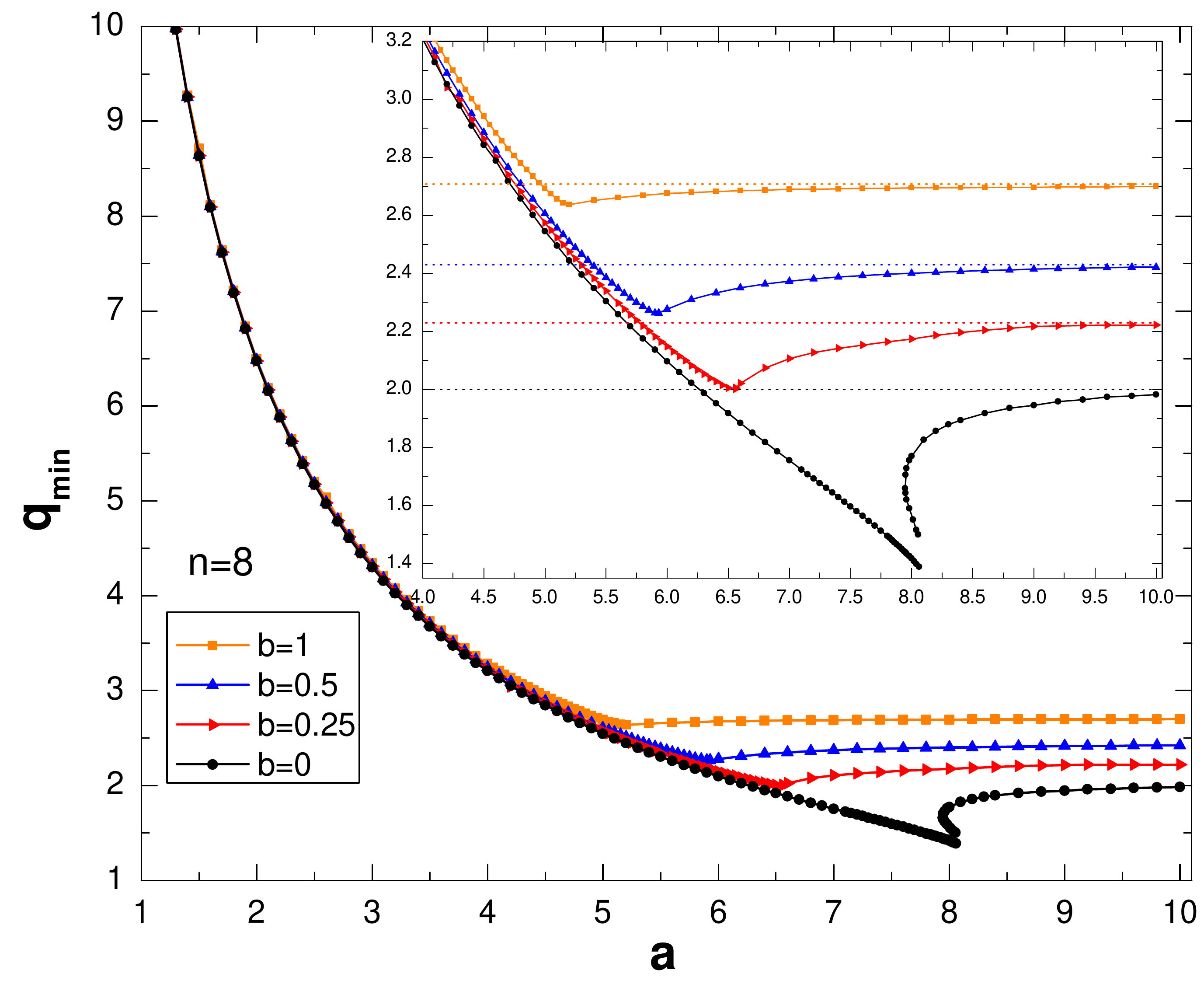}
	\end{center}
\caption{The full phase diagram of $q_{min}$ against $a$ with the Born-Infeld parameters $b =0.25,~0.5,~1$, represented by the  red, blue and orange lines, respectively.
The black line corresponds to the  case of the Maxwell model.}
		\label{qmin}
\end{figure}

To study the properties of the bound on charge-to-mass ratio, in  Fig. \ref{qmin}
we exhibit the phase diagram of  the minimal charge $q_{min}$  versus the amplitude $a$  with the Born-Infeld parameter $b =0.25,~0.5,~1$, represented by the  red, blue and orange lines, respectively.
The black line represents the  curve in the Maxwell model which has been discussed in  \cite{Crisford:2017gsb}. Comparing with the curve of  $q_{\min}-a$    curves, we can see that  the minimum value  of $q_{min}$ increases  with the increasing of the  Born-Infeld parameter $b$, Moreover, the asymptotical value $q_{min}$ of the curve appears to approach some  bound $q_w$ as $a$ increases. More details  are  shown in the  inset, and   the horizonal  black,   red, blue and orange  dashed gridlines indicate the values of  $q_w=2, ~2.23, ~2.43, ~2.708$, respectively.
It is noteworthy that for the curve of Maxwell model  in  the solutions with $\Phi\neq0$,
the critical $q_{min}$  decreases firstly with the decreasing of the  amplitude  $a$, and then it reaches a minimal point.
Further increasing $a$ to $a_{max}$,  the value of $q_{min}$  continues to decrease,
and  a second branch  with lower $q_{min}$ is obtained.
When the curve with lower $q_{min}$ moves toward $a_{max}$, the numerical error begins to increase and  a finer mesh is required to calculate.
 Though the curve of  solutions with $\Phi=0$ should connect with  the curve with $\Phi\neq0$,   it is difficult to handle numerical calculation near the value of $a_{max}$. However, for larger value of  $b$, it is relatively easy to obtain the connected curve  near the value of $a_{max}$.

\section{\label{conclusions}Conclusions}

In this paper,  we
have presented the static solutions of Einstein gravity coupled with Born-Infeld  electrodynamic and a free, massive  scalar field in four-dimensional AdS  spacetime.
We find there exists a critical value of  charge $q_{min}$, below which the static solution with zero scalar condensation is stability. However,
 when the charge of the scalar field was  above  $q_{min}$,
the static solution of Born-Infeld action will
become unstable and a new stable solution with nontrivial charged scalar field could appear, which is   analogous to the instability of a charged static solution to develop
scalar condensation in the case of Maxwell theory \cite{Crisford:2017gsb}.
Meanwhile,
there exists  the asymptotical value
of charge $q_{w}$ required to  remove the counterexamples and preserve cosmic censorship.
 Comparing with the Einstein-Maxwell model, the bound of charge-to-mass ratio $q/m$ is larger than  $1$  and increases with the Born-Infeld parameter $b$.

In the case of Maxwell theory \cite{Crisford:2017gsb}, it is surprising to find that the minimum value of the charge to mass ratio which is necessary to preserve weak cosmic censorship
 is precisely equal to the weak gravity bound.  For the nonlinear Born-Infeld theroy, we also obtain the the minimum value of the charge to mass ratio necessary to preserve weak cosmic censorship, but, because now the weak gravity conjecture in Born-Infeld electrodynamics is still not clear,
one could not determine whether the minimal value of charge-to-mass ratio is equal to  weak gravity conjecture. However,  it is sure that  a   class of  static solution with charged scalar condensate could prevent the violation of the weak cosmic censorship, and the bound of charge-to-mass ratio could
also be obtained in the case of Born-Infeld electrodynamic.

It will be  interesting to investigate the
 several further
researches.
First, since we have studied  the static solution with a free scalar condensate in Born-Infeld electrodynamic,
 we would like to investigate how  self-interactions of the scalar field to  prevent the violation of the weak cosmic censorship and the bound of charge-to-mass ratio.
The second  extension of our study is to  construct generalized multi-scalar hair configurations, such as two coexisting states of the charge scalar field
presented in Born-Infeld electrodynamic.
Finally, we are planning to study the model of
the Einstein-Born-Infeld-vector model and construct the static solution with charged vector hair necessary to preserve weak cosmic censorship in future work.

\section*{Acknowledgement}
We would like to thank  Yu-Xiao Liu, Jie Yang and Li Zhao   for  helpful discussion.  YQW would  like to thank J. Santos for discussions and correspondence about the numerical results.
Some computations were performed on the   shared memory system at  institute of computational physics and complex systems in Lanzhou university. This work was supported by the Fundamental Research Funds for the Central Universities (Grants No. lzujbky-2017-182).


\begin{thebibliography}{99}

\bibitem{Penrose:1969pc}
  R.~Penrose,
  ``Gravitational collapse: The role of general relativity,''
  Riv.\ Nuovo Cim.\  {\bf 1}, 252 (1969)
  [Gen.\ Rel.\ Grav.\  {\bf 34}, 1141 (2002)].




\bibitem{Hawking:1976ra}
  S.~W.~Hawking,
  ``Breakdown of Predictability in Gravitational Collapse,''
  Phys.\ Rev.\ D {\bf 14}, 2460 (1976).
  doi:10.1103/PhysRevD.14.2460


\bibitem{Joshi:2012mk}
  P.~S.~Joshi and D.~Malafarina,
  ``Recent developments in gravitational collapse and spacetime singularities,''
  Int.\ J.\ Mod.\ Phys.\ D {\bf 20}, 2641 (2011)
  doi:10.1142/S0218271811020792
  [arXiv:1201.3660 [gr-qc]].



\bibitem{Gregory:1993vy}
  R.~Gregory and R.~Laflamme,
  ``Black strings and p-branes are unstable,''
  Phys.\ Rev.\ Lett.\  {\bf 70}, 2837 (1993)
  doi:10.1103/PhysRevLett.70.2837
  [hep-th/9301052].

\bibitem{Hubeny:2002xn}
  V.~E.~Hubeny and M.~Rangamani,
  ``Unstable horizons,''
  JHEP {\bf 0205}, 027 (2002)
  doi:10.1088/1126-6708/2002/05/027
  [hep-th/0202189].

\bibitem{Lehner:2010pn}
  L.~Lehner and F.~Pretorius,
  ``Black Strings, Low Viscosity Fluids, and Violation of Cosmic Censorship,''
  Phys.\ Rev.\ Lett.\  {\bf 105}, 101102 (2010)
  doi:10.1103/PhysRevLett.105.101102
  [arXiv:1006.5960 [hep-th]].


\bibitem{Santos:2015iua}
  J.~E.~Santos and B.~Way,
  ``Neutral Black Rings in Five Dimensions are Unstable,''
  Phys.\ Rev.\ Lett.\  {\bf 114}, 221101 (2015)
  doi:10.1103/PhysRevLett.114.221101
  [arXiv:1503.00721 [hep-th]].


\bibitem{Figueras:2015hkb}
  P.~Figueras, M.~Kunesch and S.~Tunyasuvunakool,
  ``End Point of Black Ring Instabilities and the Weak Cosmic Censorship Conjecture,''
  Phys.\ Rev.\ Lett.\  {\bf 116}, no. 7, 071102 (2016)
  doi:10.1103/PhysRevLett.116.071102
  [arXiv:1512.04532 [hep-th]].


\bibitem{Figueras:2017zwa}
  P.~Figueras, M.~Kunesch, L.~Lehner and S.~Tunyasuvunakool,
  ``End Point of the Ultraspinning Instability and Violation of Cosmic Censorship,''
  Phys.\ Rev.\ Lett.\  {\bf 118}, no. 15, 151103 (2017)
  doi:10.1103/PhysRevLett.118.151103
  [arXiv:1702.01755 [hep-th]].


\bibitem{Horowitz:2016ezu}
  G.~T.~Horowitz, J.~E.~Santos and B.~Way,
  ``Evidence for an Electrifying Violation of Cosmic Censorship,''
  Class.\ Quant.\ Grav.\  {\bf 33}, no. 19, 195007 (2016)
  doi:10.1088/0264-9381/33/19/195007
  [arXiv:1604.06465 [hep-th]].

\bibitem{Crisford:2017zpi}
  T.~Crisford and J.~E.~Santos,
  ``Violating the Weak Cosmic Censorship Conjecture in Four-Dimensional Anti¨Cde Sitter Space,''
  Phys.\ Rev.\ Lett.\  {\bf 118}, no. 18, 181101 (2017)
  doi:10.1103/PhysRevLett.118.181101
  [arXiv:1702.05490 [hep-th]].


\bibitem{Crisford:2018qkz}
  T.~Crisford, G.~T.~Horowitz and J.~E.~Santos,
  ``Attempts at vacuum counterexamples to cosmic censorship in AdS,''
  JHEP {\bf 1902}, 092 (2019)
  doi:10.1007/JHEP02(2019)092
  [arXiv:1805.06469 [hep-th]].


\bibitem{Crisford:2017gsb}
  T.~Crisford, G.~T.~Horowitz and J.~E.~Santos,
  ``Testing the Weak Gravity - Cosmic Censorship Connection,''
  Phys.\ Rev.\ D {\bf 97} (2018) no.6,  066005
  doi:10.1103/PhysRevD.97.066005
  [arXiv:1709.07880 [hep-th]].


\bibitem{Gubser:2008px}
  S.~S.~Gubser,
  ``Breaking an Abelian gauge symmetry near a black hole horizon,''
  Phys.\ Rev.\ D {\bf 78}, 065034 (2008)
  doi:10.1103/PhysRevD.78.065034
  [arXiv:0801.2977 [hep-th]].



\bibitem{Hartnoll:2008vx}
  S.~A.~Hartnoll, C.~P.~Herzog and G.~T.~Horowitz,
  ``Building a Holographic Superconductor,''
  Phys.\ Rev.\ Lett.\  {\bf 101}, 031601 (2008)
  doi:10.1103/PhysRevLett.101.031601
  [arXiv:0803.3295 [hep-th]].

\bibitem{Cai:2015cya}
  R.~G.~Cai, L.~Li, L.~F.~Li and R.~Q.~Yang,
  ``Introduction to Holographic Superconductor Models,''
  Sci.\ China Phys.\ Mech.\ Astron.\  {\bf 58}, no. 6, 060401 (2015)
  doi:10.1007/s11433-015-5676-5
  [arXiv:1502.00437 [hep-th]].


\bibitem{ArkaniHamed:2006dz}
  N.~Arkani-Hamed, L.~Motl, A.~Nicolis and C.~Vafa,
  ``The String landscape, black holes and gravity as the weakest force,''
  JHEP {\bf 0706}, 060 (2007)
  doi:10.1088/1126-6708/2007/06/060
  [hep-th/0601001].


\bibitem{Horowitz:2019eum}
  G.~T.~Horowitz and J.~E.~Santos,
  ``Further evidence for the weak gravity - cosmic censorship connection,''
  arXiv:1901.11096 [hep-th].


\bibitem {IN-Born:1933gh}M.~Born, \textquotedblleft Modified field equations with a finite radius of the electron, \textquotedblright\
 Nature \textbf{132}, 282-282 (1933).

\bibitem {IN-Born:193444}M.~Born, \textquotedblleft Quantum theory of the electromagnetic field,\textquotedblright\ Proc.\ R.\ Soc.\ A  \textbf{143}, 410-37  (1934).


\bibitem {IN-Born:1934gh}M.~Born and L.~Infeld, \textquotedblleft Foundations
of the new field theory,\textquotedblright\ Proc.\ Roy.\ Soc.\ Lond.\ A
\textbf{144}, 425 (1934). doi:10.1098/rspa.1934.0059


\bibitem{Fradkin:1985qd}
  E.~S.~Fradkin and A.~A.~Tseytlin,
  ``Nonlinear Electrodynamics from Quantized Strings,''
  Phys.\ Lett.\  {\bf 163B}, 123 (1985).
  doi:10.1016/0370-2693(85)90205-9
\bibitem{Leigh:1989jq}
  R.~G.~Leigh,
  ``Dirac-Born-Infeld Action from Dirichlet Sigma Model,''
  Mod.\ Phys.\ Lett.\ A {\bf 4}, 2767 (1989).
  doi:10.1142/S0217732389003099


\bibitem{gsp} A.~Garc\'ia, H.~Salazar and J.F.~Pleb\'anski, ``Type-D solutions of the Einstein and Born-Infeld nonlinear-electrodynamics equations,'' Nuovo Cimento B 84, 65 (1984).


\bibitem{Cataldo:1999wr}
  M.~Cataldo and A.~Garcia,
  ``Three dimensional black hole coupled to the Born-Infeld electrodynamics,''
  Phys.\ Lett.\ B {\bf 456} (1999) 28
  doi:10.1016/S0370-2693(99)00441-4
  [hep-th/9903257].



\bibitem {IN-Fernando:2003tz}S.~Fernando and D.~Krug, ``Charged black hole
solutions in Einstein-Born-Infeld gravity with a cosmological constant,''
Gen.\ Rel.\ Grav.\ \textbf{35}, 129 (2003) doi:10.1023/A:1021315214180 [hep-th/0306120].

\bibitem {IN-Dey:2004yt}T.~K.~Dey, \textquotedblleft Born-Infeld black holes
in the presence of a cosmological constant,\textquotedblright\ Phys.\ Lett.\ B
\textbf{595}, 484 (2004) doi:10.1016/j.physletb.2004.06.047 [hep-th/0406169].

\bibitem {IN-Cai:2004eh}R.~G.~Cai, D.~W.~Pang and A.~Wang, \textquotedblleft
Born-Infeld black holes in (A)dS spaces,\textquotedblright\ Phys.\ Rev.\ D
\textbf{70}, 124034 (2004) doi:10.1103/PhysRevD.70.124034 [hep-th/0410158].




\bibitem {IN-Li:2016nll}S.~Li, H.~Lu and H.~Wei, ``Dyonic (A)dS Black Holes in
Einstein-Born-Infeld Theory in Diverse Dimensions,'' JHEP \textbf{1607}, 004
(2016) doi:10.1007/JHEP07(2016)004 [arXiv:1606.02733 [hep-th]].




\bibitem {B_F} P. Breitenlohner and D. Z. Freedman, ``Stability In Gauged Extended Supergravity,''
Annals Phys.  \textbf{144}, 249 (1982).


\bibitem{Horowitz:2014gva}
  G.~T.~Horowitz, N.~Iqbal, J.~E.~Santos and B.~Way,
  ``Hovering Black Holes from Charged Defects,''
  Class.\ Quant.\ Grav.\  {\bf 32}, 105001 (2015)
  doi:10.1088/0264-9381/32/10/105001
  [arXiv:1412.1830 [hep-th]].

\bibitem {hugeome}
 M. Abramowitz, I. A. Stegun, {\em Handbook of Mathematical Functions}, (Dover, New York,
1972).

\bibitem{Basu:2008st}
  P.~Basu, A.~Mukherjee and H.~H.~Shieh,
  ``Supercurrent: Vector Hair for an AdS Black Hole,''
  Phys.\ Rev.\ D {\bf 79}, 045010 (2009)
  doi:10.1103/PhysRevD.79.045010
  [arXiv:0809.4494 [hep-th]].

\bibitem{Hartnoll:2008kx}
  S.~A.~Hartnoll, C.~P.~Herzog and G.~T.~Horowitz,
  ``Holographic Superconductors,''
  JHEP {\bf 0812}, 015 (2008)
  doi:10.1088/1126-6708/2008/12/015
  [arXiv:0810.1563 [hep-th]].


\bibitem{Horowitz:2013jaa}
  G.~T.~Horowitz and J.~E.~Santos,
  ``General Relativity and the Cuprates,''
  JHEP {\bf 1306}, 087 (2013)
  doi:10.1007/JHEP06(2013)087
  [arXiv:1302.6586 [hep-th]].






\end{thebibliography}

\end{document}